\def\preprint{1}			% Use for submitted manuscript
\ifdefined\preprint
  \documentclass[preprint,review,12pt]{elsarticle}
\fi
\ifdefined\wordcount
  \documentclass[final,3p,times,twocolumn]{elsarticle}
\fi
\ifdefined\final
  \documentclass[final,3p,times,twocolumn]{elsarticle}
\fi

\usepackage{graphicx,stfloats}
\usepackage{color}

\usepackage{latexsym}
\usepackage{subfigure}
\usepackage{float}
\usepackage{multirow}
\usepackage{threeparttable}
\usepackage{amssymb}
\usepackage{amsthm}
\usepackage{amsmath} % Aliou 
\usepackage{tikz} % Aliou To add tex on figures 
\usepackage{notoccite} % Aliou 
\usetikzlibrary{shapes.geometric, arrows}
\usetikzlibrary{datavisualization.formats.functions}
\usetikzlibrary{positioning}
\usepackage{url}
\usepackage{mathtools}
\usepackage{pgfplots}
\usepackage{subfig}
\usepackage{float}
\usepackage{booktabs}

\newcommand{\pde}[2]{\frac{\partial {#1}}{\partial {#2}}}   % partial derivative

\newcommand{\ode}[2]{\dfrac{\text{d} {#1}}{\text{d} {#2}}}  % ordinary derivative

\biboptions{sort&compress}

\journal{Proceedings of the Combustion Institute}

\begin{document}

\begin{frontmatter}

\title{Effect of Boundary Layer Losses on 2D Detonation Cellular Structures}

\author[fir]{Qiang Xiao\corref{cor1}}
\ead{qxiao067@uottawa.ca}
\author[fir]{Aliou Sow}
\author[sec]{Brian Maxwell}
\author[fir]{Matei I. Radulescu}
%\ead{matei@uottawa.ca}
\address[fir]{Department of Mechanical Engineering, University of Ottawa, 161 Louis Pasteur, Ottawa, ON K1N6N5,Canada}
\address[sec]{Department of Mechanical and Aerospace Engineering, Case Western Reserve University, 10900 Euclid Avenue, Cleveland, OH 44106, USA}
\cortext[cor1]{Corresponding author:}

\begin{abstract}
We evaluate the effect of boundary layer losses on two-dimensional H2/O2/Ar cellular detonations obtained in narrow channels.  The experiments provide the details of the cellular structure and the detonation speed deficits from the ideal CJ speed.  We model the effect of the boundary layer losses by incorporating the flow divergence in the third dimension due to the negative boundary layer displacement thickness, modeled using Mirels' theory.  The cellular structures obtained numerically with the resulting quasi-2D formulation of the reactive Euler equations with two-step chain-branching chemistry are found in excellent agreement with experiment, both in terms of cell dynamics and velocity deficits, provided the boundary layer constant of Mirels is modified by a factor of 2.   A significant increase in the cell size is found with increasing velocity deficit.  This is found to be very well captured by the induction zone increase in slower detonations due to the lower temperatures in the induction zone.   
\end{abstract}


\end{frontmatter}

%% Please do not modify the following three lines
\ifdefined \wordcount
\clearpage
\fi

\section{Introduction}
\label{Introduction}

\textcolor{black}{Detonation structures in narrow channels are usually observed to exhibit a two-dimensional (2D) structure (\textcolor{black}{e.g., see Refs. \cite{austin2003, bhattacharjee2013, xiao2019}}).}  Nevertheless, they \textcolor{black}{tend to} propagate with large velocity deficits due to losses \textcolor{black}{originating from} the side walls \cite{berets1950, manson1957}. The cell sizes are also reported to be much larger in \textcolor{black}{narrower} channels, due presumably to increased reaction zone lengths \cite{strehlow1967, monwar2007, ishii2011}, \textcolor{black}{and the propagation limits (i.e., the critical initial pressures, below which detonations fail to propagate) are impacted as well (e.g., see Refs. \cite{dove1974, radulescu2002failure,chao2009,zhang2016b}).}

Previous works focused on \textcolor{black}{modelling the wall losses of detonations in 1D \cite{zeldovich1950, fay1959, zhang1994, sow2014, faria2015}}.  One approach, introduced by Zel'dovich \cite{zeldovich1950},  \textcolor{black}{was} to model the wall losses with volume-averaged friction and heat loss terms.  Another, \textcolor{black}{due to Fay \cite{fay1959}}, \textcolor{black}{accounted} for the negative displacement thickness of the boundary layer, whose effect \textcolor{black}{appeared} as a source term of \textcolor{black}{mass} flow sink in the governing equations for the inviscid core \textcolor{black}{(i.e., the undisturbed free stream flow)}. Despite these efforts in modelling 1D detonations, in a realistic way, the effect of these wall losses on the dynamics of 2D cellular detonations in narrow channels remains unknown. 

\textcolor{black}{On the other hand, the recent works of Tsuboi et al. \cite{tsuboi2013}, Chinnayya et al. \cite{chinnayya2013}, and Sow et al. \cite{sow2019} have showed that directly resolving the viscous boundary layers by the Navier-Stokes (NS) equations requires remarkably high resolution, which makes the computations considerably expensive. Moreover, for these NS calculations, the other difficulty lies in quantifying the wall loss effects on detonation cellular structures.}

In the present study, we adapt Mirels'  technique \textcolor{black}{\cite{mirels1956}} by accounting for the wall-boundary-layer-induced loss in a 2D formulation of the problem.  This permits \textcolor{black}{us} to \textcolor{black}{readily} compute the dynamics of unsteady 2D cellular detonations with a supplemental lateral loss. 

The communication first reports experiments in a narrow channel, with variation of the channel width ($w$) and cell size ($\lambda$) ratios, i.e., $w/\lambda$.  We then formulate the \textcolor{black}{governing} equations with a lateral loss, which is evaluated from Mirels' boundary layer theory \cite{mirels1956}. Comparisons between experiments and simulations follow. Finally, \textcolor{black}{effects of the boundary layer losses on dynamics of the unsteady 2D cellular detonations are explored}. 

\section{Experiments in a Narrow Channel}
\subsection{Experimental details}
\label{Experimental details}
The experiments were performed in a 3.4-m-long thin rectangular aluminium channel with an internal height and width of 203 mm and 19 mm, respectively, as described in detail elsewhere \cite{bhattacharjee2013}. The shock tube comprises three parts, i.e., the detonation initiation section, the propagation section, and the test section (about 1.0 m in length). The mixture was ignited in the first section by a high voltage igniter, and mesh wires were inserted in this part for promoting the detonation formation.  The detonation evolution process was visualized in the test section, and its mean propagation speed over the whole test part was obtained by \textcolor{black}{one 113B24 and five 113B27 piezoelectric PCB pressure sensors} using the time-of-arrival method. The presently investigated mixture is the very regular stoichiometric hydrogen-oxygen diluted with $70\%$ argon (i.e., 2H$_2$/O$_2$/7Ar). Since the mixture of 2H$_2$/O$_2$/7Ar has low reactive sensitivity, a more reactive driver gas of stoichiometric ethylene-oxygen (i.e., C$_2$H$_4$/3O$_2$) was used in the initiation section, which was separated from the propagation section with a diaphragm. For visualization, a Z-type schlieren setup \cite{bhattacharjee2013} was utilized with a light source of 360 W. The resolution of the high-speed camera was $384\times288$ px$^2$ with the framing rate of 77481 fps (about 12.9 $\mu s$ for each interval).   
\subsection{Results}
 \label{Results}
\begin{figure}[]
 	\centering
 	{\includegraphics[width=1.0\textwidth]{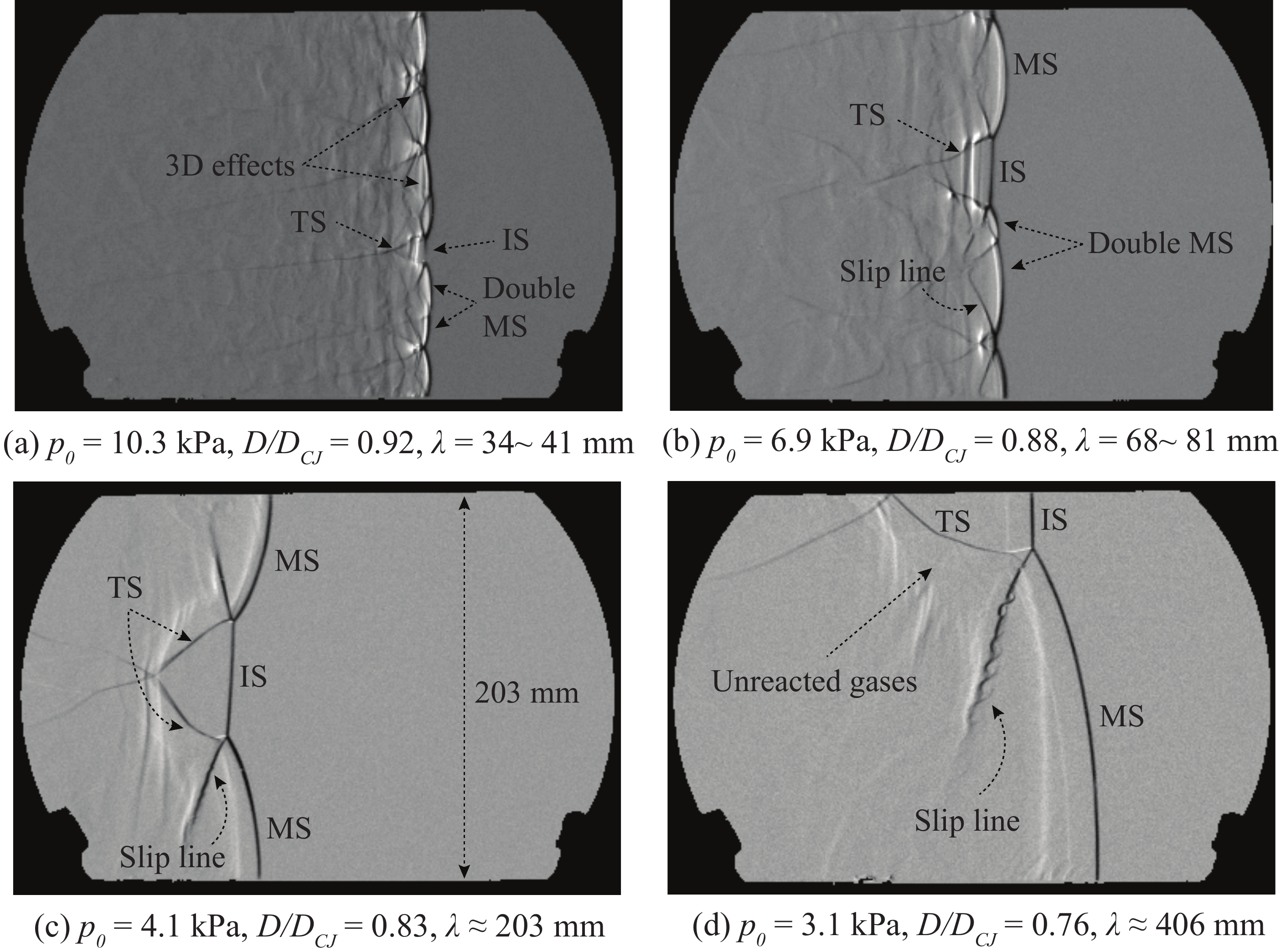}}
 	\caption{\textcolor{black}{The structure of cellular detonations with varied $w/\lambda$, IS: the incident shock, MS: Mach stem, TS: the transverse shock (video animations as Supplemental material illustrating the evolution process).} } \label{ExpPhoto}  
 \end{figure}

Figure \ref{ExpPhoto} shows the schlieren photos of the detonation reaction zone structures, at varied initial pressures ranging from 10.3 kPa to 3.1 kPa. By decreasing the initial pressure for reducing the kinetic sensitivity of the mixture, detonations can be clearly observed to propagate with considerably enlarged cellular structures, with the velocity deficits increased up to $20\% \sim 30\%$ of the ideal CJ detonation speed. At a relatively high initial pressure in Fig.\ \ref{ExpPhoto}a, where $w/\lambda$ is about 0.5, \textcolor{black}{one can observe possible 3D-like effects of the overall detonation structure by the presence of duplicate features that do not overlap in the schlieren image.} This adds to the difficulty in studying its dynamics under the present resolution. With the decrease of $w/\lambda$ to about 0.25 in Fig.\ \ref{ExpPhoto}b, the detonation structure is qualitatively similar. The appearance of double Mach stems sharing the same triple point indicates that detonation is still relatively unstable in this condition. When $w/\lambda$ is further reduced to less than 0.1, as shown in Fig.\ \ref{ExpPhoto}c and Fig.\ \ref{ExpPhoto}d, detonations become perfectly planar and \textcolor{black}{essentially} two-dimensional for the investigation.  At these low initial pressures, detonations are organized with relatively large unburned induction zones, as can be observed behind both the leading shock and the transverse wave. The vortex structures characteristic of the Kelvin-Helmholtz instability can also be readily seen along the slip line from Fig.\ \ref{ExpPhoto}d.  Since the single-head detonation in Fig.\ \ref{ExpPhoto}d experiences a much larger velocity deficit, being more expensive in CFD calculations, we will adopt the condition of Fig.\ \ref{ExpPhoto}c as the main benchmark for subsequent model development and validation of simulations.
 
\section{Numerical Simulations}
\label{Numerical Simulations}
 \subsection{Unsteady quasi-2D formulation}
 \label{Unsteady quasi-2D formulation}
 Although the viscous effects are responsible for the boundary layer losses of detonations in thin channels, the present work aims at modelling these effects into the inviscid core flow as a source term. For such transient inviscid reactive core flow behind detonations in a narrow channel, the area-averaged equations of motion across the $z$-direction (i.e., the channel width direction into/out of the page when viewing Fig.\ \ref{ExpPhoto}) can be expressed in the lab frame of reference $\left(x, y, t\right)$ as 
 \begin{subequations}
 	\begin{align}
 	&\pde{\rho}{t}+\pde{\left(\rho u\right)}{x} + \pde{\left(\rho v\right)}{y} = -\rho\dfrac{1}{A}\dfrac{DA}{Dt} \label{mass-2} \\
 	&\pde{\left(\rho u\right)}{t}+\pde{\left(\rho u^2 +p\right)}{x}+\pde{\left(\rho u v\right)}{y} = -\rho u\dfrac{1}{A}\dfrac{DA}{Dt} \label{momentum-x-2} \\
 	&\pde{\left(\rho v \right)}{t}+\pde{\left(\rho u v \right)}{x}+\pde{\left(\rho v^2 +p\right)}{y} = -\rho v\dfrac{1}{A}\dfrac{DA}{Dt} \label{momentum-y-2} \\
 	&\textcolor{black}{\pde{\left(\rho e\right)}{t} + \pde{\left(\rho e u +p u\right)}{x} + \pde{\left(\rho e v +p v\right)}{y} = -\left(\rho e+p\right) \dfrac{1}{A}\dfrac{DA}{Dt} -Q\dot{\omega}_R \label{energy-2}} \\
 	&\textcolor{black}{\pde{\left(\rho Y \right)}{t} + \pde{\left(\rho uY\right)}{x} +\pde{\left(\rho v Y\right)}{y} = -\rho Y \dfrac{1}{A}\dfrac{DA}{Dt} +\dot{\omega}_R} \label{species-2}
 	\end{align} \label{governing-equations-2}where $\rho, u, v, A, p,\textcolor{black}{Q, Y, \dot{\omega}_R}$ denote the mixture density, $x$-direction flow velocity, $y$-direction flow velocity, the cross section area, pressure, \textcolor{black}{heat release, mass fraction and rate of mass production of the single reactant $R$. Note that the present work adopts the reaction of single species with no intermediate radicals, i.e., in the form of $R \rightarrow P$, where $R$ is the only reactant and $P$ the product. The total sensible energy plus the kinetic energy is $e = \dfrac{p/\rho}{\gamma - 1}+\dfrac{1}{2}\left(u^2+v^2\right)$, where a calorically  perfect gas with constant specific heats is assumed and  $\gamma$ is the ratio of specific heats.} The material derivative $DA/Dt$ appearing in the right-hand source term is the cross-sectional area change.  Of noteworthy is that one can find the formal derivation of the 1D version  by Chesser \cite{chesser1992}.
 \end{subequations}

For an observer travelling in the frame of reference attached to the leading shock at average speed $D_s$ and following the motion of the fluid, we have the source term in Eq.\ \ref{governing-equations-2} further expressed as
\begin{align}
\dfrac{D}{Dt}\left( \ln A\right) = \pde{}{t'}\left( \ln A\right) + u'\pde{}{x'}\left( \ln A\right)
\end{align}
where $A\left(x, t\right) = H\times W\left(x, t\right)$ is the effective cross section area. $H$ is the fixed channel height of 203 mm in this communication, while $ W\left(x, t\right)$ is the effective channel width in the $z$-direction. \textcolor{black}{Since the channel height $H$ is much larger than the width $w$ (i.e., $H/w \approx 10$) in the present work, the boundary layer effects in the channel height direction are negligible as compared to those from the channel width direction.}  $x'$, $t'$, and $u'$ are, respectively, the space and time coordinates and the post-shock flow velocity in the shock-attached frame of reference. Since the motion is pseudo-steady (travelling wave with perturbation), we can neglect $\pde{}{t'}\left(lnA\right)$ to the leading order. We thus get
\begin{align}
\dfrac{D}{Dt}\left( \ln A\right) \approxeq u'\pde{}{x'}\left( \ln A \right) \label{DlnA/Dt}
\end{align} 
which can be evaluated from Fay's boundary layer theory \cite{fay1959} by using Mirels' compressible laminar boundary layer solutions \cite{mirels1956}. The boundary layer displacement thickness $\delta^*\left(x'\right)$ behind a moving shock is \cite{mirels1956}
\begin{align}
\delta^*\left(x'\right) = K_M \sqrt{\dfrac{\mu_s x'}{\rho_0 D_s}}
\end{align}
where $x'$ is the distance from the shock, $\rho_0$ the density of the flow ahead of the shock, $\mu_s$ the post-shock dynamic viscosity, and $K_M$ the Mirels' constant. One can refer to the recent work of Xiao and Radulescu \cite{xiao2019} for details in evaluating $K_M$ for hydrogen-oxygen-argon detonations at varied initial pressures. For 2H$_2$/O$_2$/7Ar detonations, they found that $K_M \approx 4.0$ \cite{xiao2019}. Using this relation, the boundary layer displacement thickness $\delta^*$  in the experiment of Fig.\ \ref{ExpPhoto}c was calculated to be 1.9 mm, with $D_s = 0.8 D_{CJ}$ and $x'$ as the hydrodynamic thickness $x_H$ between the leading shock and the sonic surface. Note that $x_H$ was computed from the generalized ZND model with lateral losses \cite{klein1995, xiao2019}. Clearly, the boundary layer is much thinner than the channel width.

Since $ W\left(x'\right) = w + 2\delta^*\left(x'\right)$, where $w$ is the physical channel width of 19 mm, we can thus obtain
\begin{align}
\pde{}{x'}\left(lnA\right) = \dfrac{2}{w + 2\delta^*\left(x'\right)}\times\ode{\delta^*\left(x'\right)}{x'}
\end{align}
where $\delta^*\left(x'\right) \ll w$, Eq.\ \ref{DlnA/Dt} then changes to
\begin{align}
\dfrac{D}{Dt}\left( \ln A\right) = u' \dfrac{2}{w} \dfrac{K_M}{2}\sqrt{\dfrac{\mu_s }{\rho_0 D_s}}\left(x'\right)^{-0.5} \label{DlnA/Dt-2}
\end{align}
Since Mirels' model assumes that the post-shock state is uniform and steady, to the leading order, we can thus write $x'$ as $x' = u'\left(t-t_s\right)$, where $t_s$ is the time at  which the particle crosses the shock. With the mass conservation across the shock $\rho_s u' = \rho_0 D_s$, where $\rho_s$ is the post-shock density, Eq.\ \ref{DlnA/Dt-2} can be greatly simplified as the following simple expression
\begin{align}
\dfrac{D}{Dt}\left( \ln A\right) =\dfrac{K_M}{w}\sqrt{\dfrac{\nu_s}{t-t_s}} \label{final-DADT}
\end{align} 
where $\nu_s$ is the post-shock kinematic viscosity, and $t_{\textrm{elapse}} = t-t_s$ is the elapsed time since a particle has passed through the shock front. The shock time $t_s$ is recorded when the shock passes over, and convected with the motion of that particle:
\begin{align}
\pde{t_s}{t} + \vec{u}\boldsymbol{\cdot} \mathbb{\nabla}t_s = 0
\end{align}

\subsection{Two-step chemistry model}
 \label{Two-step chemistry model}
In the experiment of Fig.\ \ref{ExpPhoto}c, we have calculated the post-shock temperatures using the experimentally measured shock speeds along the walls and cell axis. We found that the lowest post-shock temperature is about 900 K, which is still above its cross-over temperature of 800 K to 850 K. Thus, in this study, the two-step chain-branching reaction model \cite{short2003, leung2010} will be employed for describing the chemical kinetics. It consists of two components, i.e., a thermally neutral induction zone followed by an exothermic main reaction zone. The transport equations of the induction and reaction variables can be written as:

\begin{subequations}
%	\begin{align}
%	\textnormal{Induction:} \qquad \dfrac{D\lambda_i}{Dt}&= - \mathcal{H}\left(\lambda_i\right)k_i \rho^{\alpha}  \textrm{exp}\left(-\dfrac{E_a}{R
%		T}\right)  \label{induction}\\
%	\textnormal{Reaction:} \qquad \dfrac{D\lambda_r}{Dt}&=-\left[1-\mathcal{H}\left(\lambda_i\right)\right]k_r \rho^{\beta}\lambda_r^{\nu}  \label{reaction}
%	\end{align} 
\begin{align}
&\pde{\left(\rho \lambda_i \right)}{t} + \pde{\left(\rho u\lambda_i\right)}{x} +\pde{\left(\rho v \lambda_i\right)}{y} =  -\rho \lambda_i \dfrac{1}{A}\dfrac{DA}{Dt} - \mathcal{H}\left(\lambda_i\right)k_i \rho^{\alpha+1}  \textrm{exp}\left(-\dfrac{E_a}{R
	T}\right)  \label{induction}\\
&\pde{\left(\rho \lambda_r \right)}{t} + \pde{\left(\rho u\lambda_r\right)}{x} +\pde{\left(\rho v \lambda_r \right)}{y} =  -\rho \lambda_r \dfrac{1}{A}\dfrac{DA}{Dt} - \left[1-\mathcal{H}\left(\lambda_i\right)\right]k_r \rho^{\beta+1}\lambda_r^{\nu} \label{reaction}
\end{align}\label{reaction-rate-equations}where $\lambda_i$ is the progress variable for the induction zone with a value of 1 in the reactants and 0 at the end of the induction zone, $\lambda_r$ \textcolor{black}{the reaction progress variable} with a value of 1 in the unburned zone and 0 in the burned products. $\mathcal{H}\left(\lambda_i\right)$ is the Heaviside function given as 
\end{subequations}

\begin{align}
	\textcolor{black}{\mathcal{H}\left(\lambda_i\right) =
		\begin{cases}
		0 & \text{if $\lambda_i=0$} \\
		1 & \text{if $\lambda_i > 0$} 
		\end{cases}}
\end{align}which disables the progress of $\lambda_i$ at the end of the induction zone.  $k_i$  and  $k_r$ are rate constants, $E_a$ the activation energy controlling the temperature sensitivity of the induction zone duration, $\nu$ is the reaction order,  while $\alpha$ and $\beta$ are further empirical reaction order parameters.

Table\ \ref{2-step-parameters} shows the non-dimensional parameters for the two-step model at three different initial pressures from experiments in Fig.\ \ref{ExpPhoto}.  They were
\begin{table}[]
	\footnotesize
	\centering
	\caption{The calibrated non-dimensional parameters for the two-step model from the detailed chemistry.}
	\begin{tabular}{cccccc}
		\toprule
		$p_0$ (kPa) & $\gamma$ & $E_a/RT_0$    & $Q/RT_0$     & $k_i$    & $k_r$ \\
		\midrule
		4.1   & 1.5   & 31.2  & 11.5  & 45.6  & 0.078 \\
		6.9   & 1.5   & 22.8  & 11.8  & 10.6  & 0.11 \\
		10.3  & 1.5   & 24.2  & 12.0    & 12.7  & 0.14 \\
		\bottomrule
	\end{tabular}%
	\label{2-step-parameters}%
\end{table}%
\begin{figure}[]
	\centering
	{\includegraphics[width=1.0\textwidth]{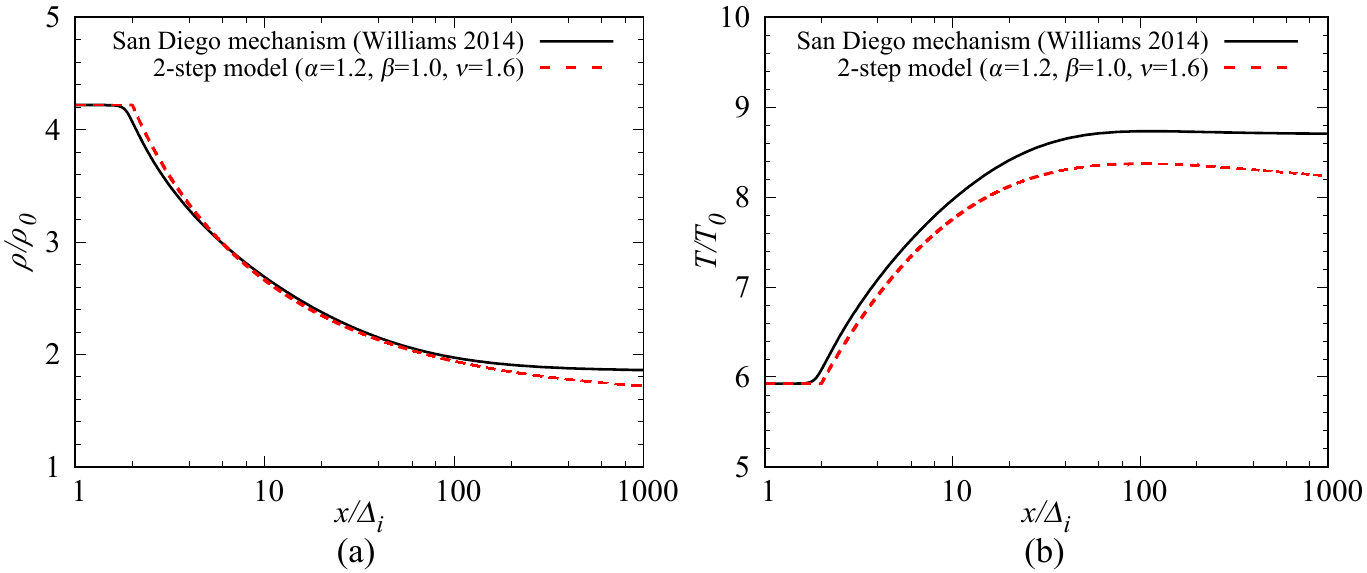}}
	\caption{\textcolor{black}{ZND profiles of (a) density and (b) temperature obtained by using the 2-step chemistry model and the detailed chemistry  at the initial pressure of 4.1 kPa.}} \label{ZND-structure} 
\end{figure}calibrated from the detailed chemistry using the San Diego chemical reaction mechanism (Williams) \cite{Williams2014}, by using Shepherd's Shock and Detonation Toolbox (SDToolbox) \cite{SDtoolbox}. $\gamma$ was the post-shock isentropic exponent of the CJ detonation, while the heat release $Q$ was determined from the perfect gas relation recovering the correct Mach number \cite{fickett2000}. The effective activation energy $E_a$ was calculated from the \textcolor{black}{logarithmic} derivative of the ignition delay with respect to the inverse of post-shock temperature. Note that the present work adopts the initial state variables ($p_0, \rho_0, T_0$) and the ZND induction zone length ($\Delta_i$) as the normalization scales. As such, the rate constant $k_i$ can be directly obtained from Eq.\ (\ref{induction}) by scaling the dimensionless induction zone length to unity, while $k_r$ can be determined by recovering the correct induction/reaction time ratio from the detailed chemistry. Finally, $\alpha = 1.2, \beta = 1.0, \nu = 1.6$ were adopted for matching the detailed chemistry ZND structure, as can be easily seen from the \textcolor{black}{density and temperature profiles} in Fig.\ \ref{ZND-structure}. \textcolor{black}{For the minor discrepancies observed near the end of the reaction zone, it is due to the limitation of assuming the constant specific heats in the present work.} 

% Table generated by Excel2LaTeX from sheet 'Sheet2'

\subsection {Computational details}
\label{Computational details}
The non-dimensional governing equations were solved employing the $MG$ code, developed by S. Falle of the University of Leeds, which uses a \textcolor{black}{second-order-accurate exact} Godunov solver \cite{falle1991} with adaptive mesh refinement. The computational domain height was held constant at the same height of experiments (203 mm in height), i.e., $72\Delta_i$ for $p_0 = 4.1$ kPa, $116\Delta_i$ for $p_0 = 6.9$ kPa, and $182\Delta_i$ for $p_0 = 10.3$ kPa. The domain length varied from $3000\Delta_i$ to $5000\Delta_i$. The detonation propagated from left to the right, with reflective boundary conditions imposed to the top and bottom sides, and zero-gradient boundary conditions applied to the left and right ends. The computations were started using a ZND profile placed $300\Delta_i$ in length from the left boundary. An initial density disturbance zone of $4\Delta_i$ was added ahead of the initial ZND solution for accelerating the evolution to cellular detonations. \textcolor{black}{This density perturbation method is the same as that of Maxwell et al. \cite{maxwell2017}, which is give by} 
\begin{align}
\textcolor{black}{\rho\left(x, y, t = 0\right) =
	\begin{cases}
	\textnormal{ZND solution} & \text{if $x < 0$} \\
	1.25 - 0.5n & \text{if $0 \le x \le 4$} \\
	1 & \text{otherwise} \\
	\end{cases}}
\end{align}\textcolor{black}{where $n$ is a random real number from 0 to 1.} As for the numerical resolution, 5 levels of mesh refinement were adopted with the coarsest and finest grid sizes of $1/2\Delta_i$ and $1/16\Delta_i$, respectively. Since the reaction zone length of the simulated cases in this work is in the order of $100\Delta_i$ to $1000\Delta_i$, as demonstrated in Fig.\ \ref{ZND-structure}, such resolution is adequate for obtaining reliable results. This has been verified by a resolution test using a higher level of mesh refinement for calculating the CJ detonation without loss, at the initial pressure of 4.1 kPa. We found that the final stable cell size does not change. In simulations, we ran all the cases for long enough time until we have obtained at least 10 repeated stable cycles of the detonation structures. Due to the large domain size, each case running with 100 to 200 cores in parallel in \textit{Cedar} of Compute Canada requires about one week to complete. More than 20 cases were involved in this study.  

\subsection {Results and discussion}
\label{results}
\subsubsection{Comparison with experiments}
\label{Comparison with experiments}
\begin{figure}[]
	\centering
	{\includegraphics[width=1.0\textwidth]{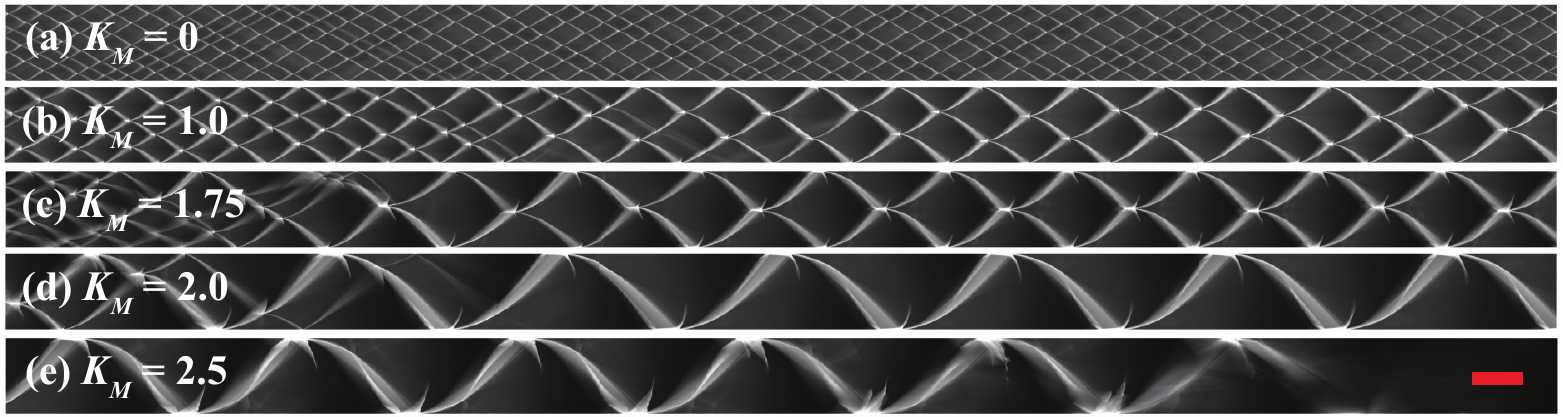}}
	\caption{The recorded maximum energy release rates of detonations at the initial pressure of 4.1 kPa with varied $K_M$. Their cell size and mean propagation speeds are: (a) $D/D_{CJ} = 1.0, \lambda =  51$ mm, (b) $D/D_{CJ} = 0.90, \lambda =  136$ mm,  (c) $D/D_{CJ} = 0.85, \lambda =  203$ mm, (d) $D/D_{CJ} = 0.81, \lambda \approx  406$ mm , and (e) detonation failure. In the experiment, $D/D_{CJ} = 0.83, \lambda =  203$ mm . Note that the red symbol represents  $50\Delta_i$ in length, and the length of the shown domain is $1500\Delta_i$.} \label{shutr-2}  
\end{figure}

\begin{figure}[]
	\centering
	{\includegraphics[width=1.0\textwidth]{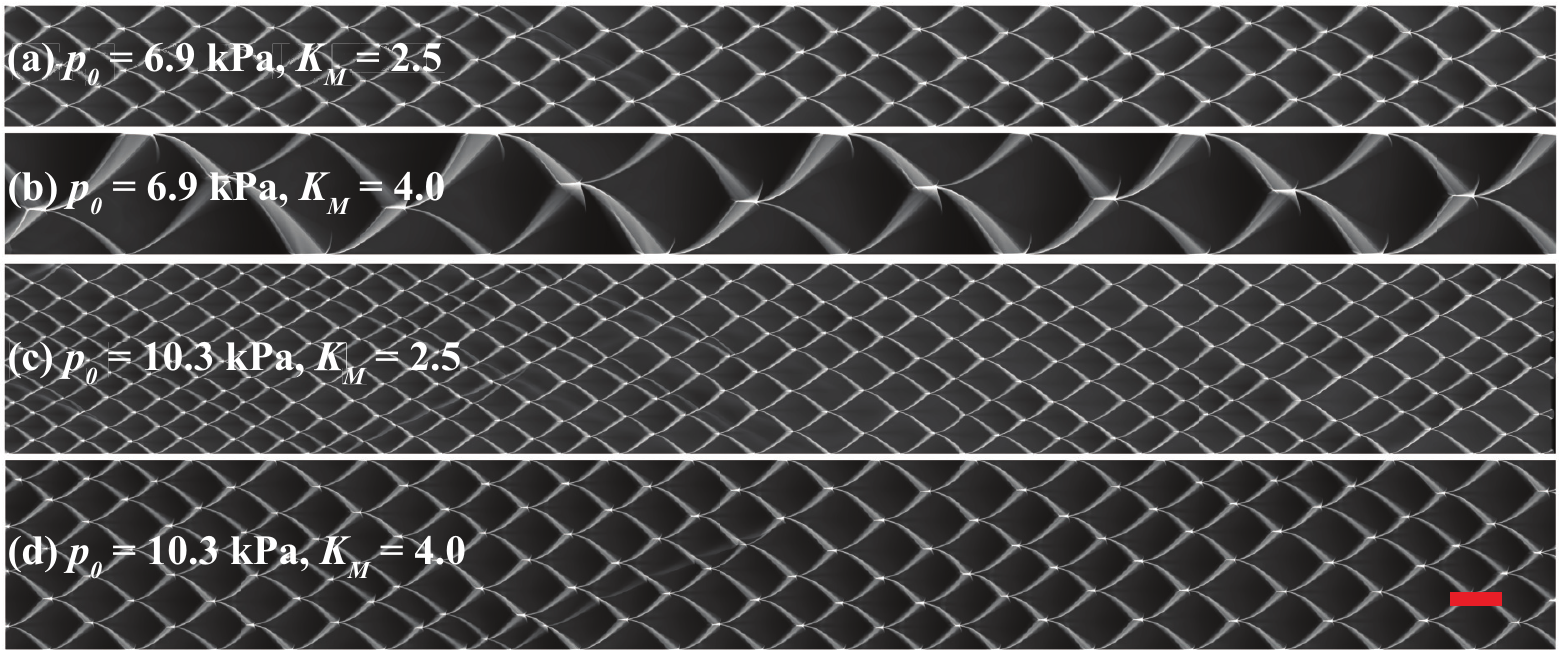}}
	\caption{The recorded maximum energy release rates of detonations at the initial pressure of 6.9 kPa and 10.3 kPa, respectively. Their cell size and mean propagation speeds are: (a) $D/D_{CJ} = 0.87, \lambda =  81$ mm, (b) $D/D_{CJ} = 0.78, \lambda =  203$ mm,  (c) $D/D_{CJ} = 0.91, \lambda =  41$ mm, (d) $D/D_{CJ} = 0.87, \lambda \approx  58$ mm.  Note that the red symbol represents  $50\Delta_i$ in length, and the length of the shown domain is $1500\Delta_i$. } \label{shutr-4}  
\end{figure}

Figure\ \ref{shutr-2} shows the numerically tracked maximum energy release rates for detonations with different losses, at the initial pressure of 4.1 kPa. This corresponds to the open shutter photograph in experiments. From the simulated results, it can be observed that the detonation cell size becomes larger as a result of increasing the Mirels' constant $K_M$, i.e., increasing the magnitude of boundary layer losses. As the ideal CJ detonation ($K_M = 0$) has four stable cells across the channel height, it can only accommodate a single-head detonation for $K_M = 2.0$. When $K_M $ is further increased to 2.5, the single-head detonation finally failed, as can be seen from Fig.\ \ref{shutr-2}e. According to the calculations performed by Xiao and Radulescu \cite{xiao2019}, the theoretical Mirels' constant for  2H$_2$/O$_2$/7Ar detonations is supposed to be $K_M \approx 4.0$. However, the present simulations show that $K_M = 1.75$ can very well recover the experiment (in Fig.\ \ref{ExpPhoto}c) in terms of the cell size and the velocity deficit. This also occurs for cases at other initial pressures of 6.9 kPa and 10.3 kPa, respectively, as shown in Fig. \ref{shutr-4}. While $K_M = 4.0$ results in a larger cell size and velocity deficit, $K_M \approx 2.5$ appears to be able to correctly recover them, when compared to the experiments from Fig.\ \ref{ExpPhoto}a and Fig.\ \ref{ExpPhoto}b. Such discrepancy from the theoretically computed $K_M$ presumably originates from Mirels' assumption of the uniform and steady state behind the shock. For detonations, significant gradients of pressure, temperature, \textcolor{black}{and velocity} exist. \textcolor{black}{Particularly, the flow acceleration (in the shock-attached reference) from being subsonic behind the detonation front to sonic in the reaction zone can contribute to thinned boundary layers, as already noted by Chinnayya et al. \cite{chinnayya2013}. In their 2D viscous simulations, they also found that the thickness of the computed boundary layers behind detonations is approximately half of that predicted by the uniform steady boundary layer theory.} Moreover, the theoretical calculations of $K_M$ assume the leading shock of the ideal CJ detonation speed, while the 2D simulations have detonations of significant velocity deficits. Thus, it results in a smaller $K_M$ than expected by theory. Future work should be devoted to refine the model to account for these non-idealities. Nevertheless, it is quite satisfying that the model works within a factor of 2, which can also be absorbed by uncertainties in chemical kinetics \cite{taylor2013}.

\begin{figure}[]
	\centering
	{\includegraphics[width=1.0\textwidth]{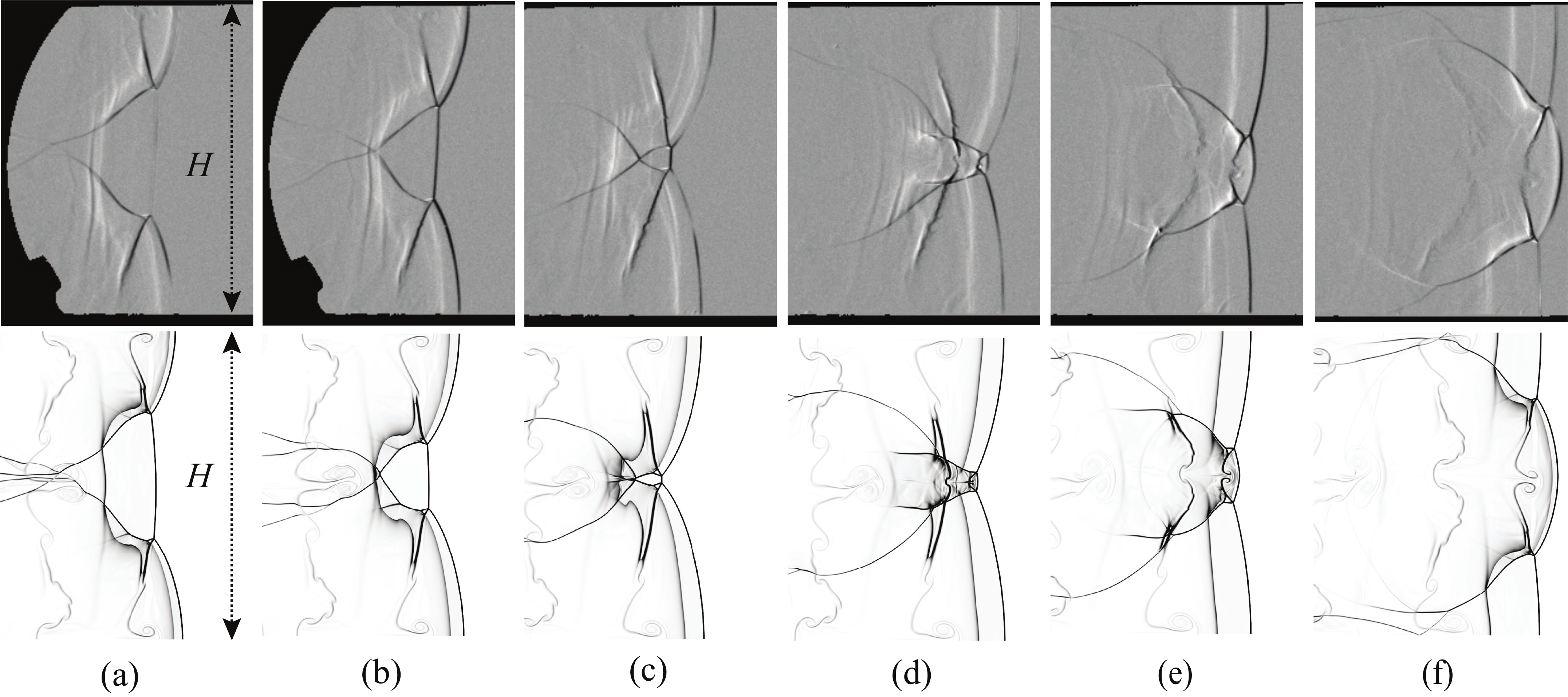}}
	\caption{Comparisons of the gradient of the density (bottom column) from the quasi-2D simulation ( $p_0 = 4.1$ kPa, $K_M = 1.75$) with the schlieren photos (top column) from experiments at the initial pressure of 4.1 kPa. $H$ is the channel height of 203 mm. } \label{Exp-compare-0.6psi}  
\end{figure}

\begin{figure}[]
	\centering
	\includegraphics[width=0.62\textwidth]{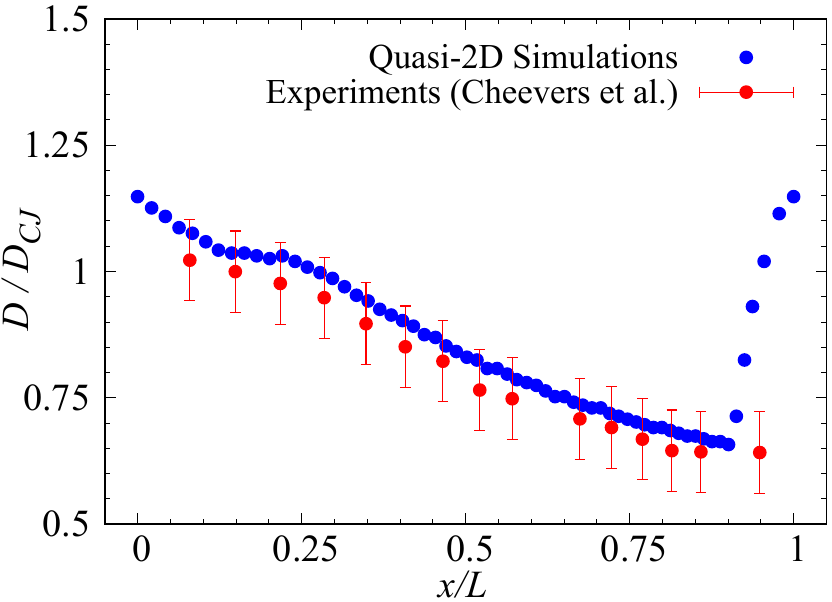}
	\caption{The evolution of detonation speeds in a cell at the initial pressure of 4.1 kPa. Note that the experimental speeds in a cell were reconstructed by Cheevers et al. \cite{cheevers2019}. The numerical data are the local time-average speed obtained with a running time average of 5 neighbouring points from more than 60 sampling points in a stable cell, for the case of $p_0 = 4.1$ kPa with $K_M = 1.75$. \textcolor{black}{$L$ is the cell length.}} \label{Dspeed-compare-0.6psi}  
\end{figure}

Besides the velocity deficit and cell size, the experimentally visualized qualitative features of the detonation structure, as well as its cellular dynamics can be very well reproduced by the simulations, as shown in Fig.\ \ref{Exp-compare-0.6psi}. Figure\ \ref{Dspeed-compare-0.6psi} also shows the quantitative agreement in temporal velocity evolution at a single cell, when compared to experiments. This suggests the robustness of the proposed quasi-2D formulation as well as the two-step chemistry model in simulating the real detonations in experiments.

\subsubsection{Dynamics of detonations with different losses}
\begin{figure}[]
	\centering
	{\includegraphics[width=0.62\textwidth]{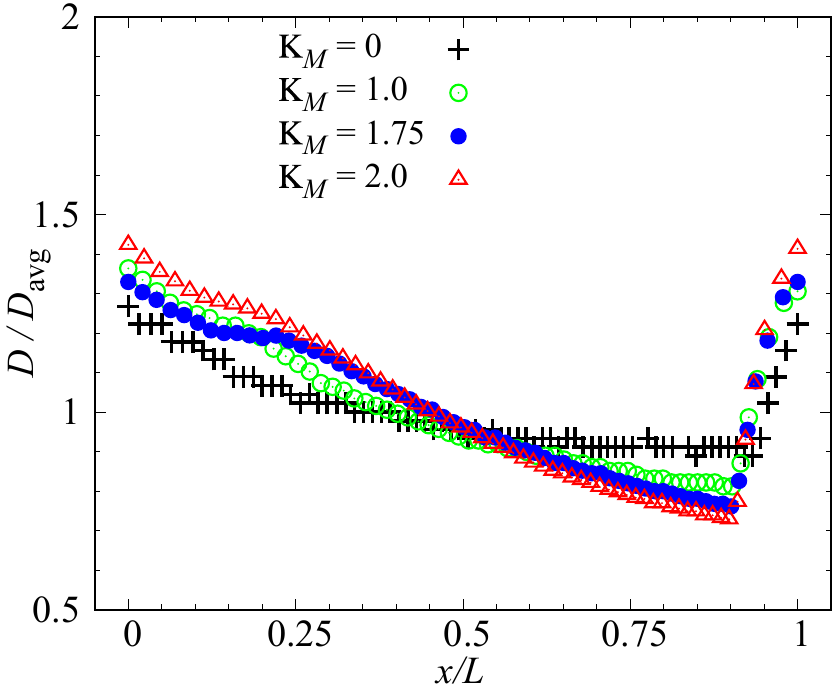}}
	\caption{Local time-average speed (along the cell axis) of detonations with different losses, at the initial pressure of 4.1 kPa. Note that the detonation speed is normalized by the mean propagation speed $D_{\textrm{avg}}$ in a cell, and $L$ is the cell length. } \label{Dspeed-Km} 
\end{figure}

The effect of boundary layer losses on detonation dynamics is shown in Fig.\ \ref{Dspeed-Km}, in terms of the normalized speed with respect to the position in a cell. More than 60 points were sampled in a cell for each case, and the local time-average speed was obtained with a running time average of 5 neighbouring points.  Evidently, the presence of wall losses modifies the cellular dynamics. Compared to the CJ case, the cases with losses have a larger deviation from the larger maximum velocity at the beginning of the cell to the smaller minimum one before the collision of triple points. Whether it is the losses that directly modify the flow velocity inside the cellular structure, or the increased activation energy due to velocity deficits that results in such larger fluctuation, requires further confirmation.

\subsubsection{The ${D}/{D_{CJ}}-K_M$ relationships}
\begin{figure}[]
	\centering
	\includegraphics[width=0.6\textwidth]{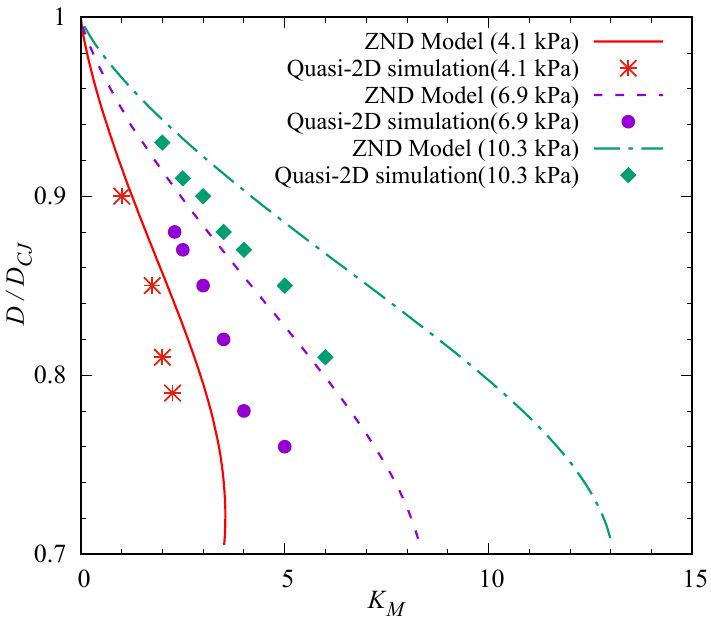}
	\caption{Comparisons of the quasi-2D simulation results with the steady 1D ZND model in terms of the  ${D}/{D_{CJ}}-K_M$ relationships.} \label{2step-ZND}  
\end{figure}

The variation of the velocity deficit with respect to the Mirels' constant $K_M$ can be better appreciated from Fig.\ \ref{2step-ZND}. The theoretical predictions were obtained by solving the generalized ZND model with lateral flow divergence \cite{klein1995}, using the present two-step reaction model. Clearly, the ZND model underpredicts the velocity deficit obtained from the quasi-2D simulations. As the constant $K_M$ increases to the propagation limit, such discrepancy becomes more significant. These results are consistent with those reported previously by Sow et al. \cite{sow2014} and Reynaud et al. \cite{reynaud2017}, who also found the underprediction of the velocity deficit from the unsteady simulations by the analytical model for relatively regular detonations.   

\subsubsection{The ${\lambda}/{\lambda_{CJ}} - {D}/{D_{CJ}}$ relationships}
\begin{figure}[]
	\centering
	{\includegraphics[width=0.6\textwidth]{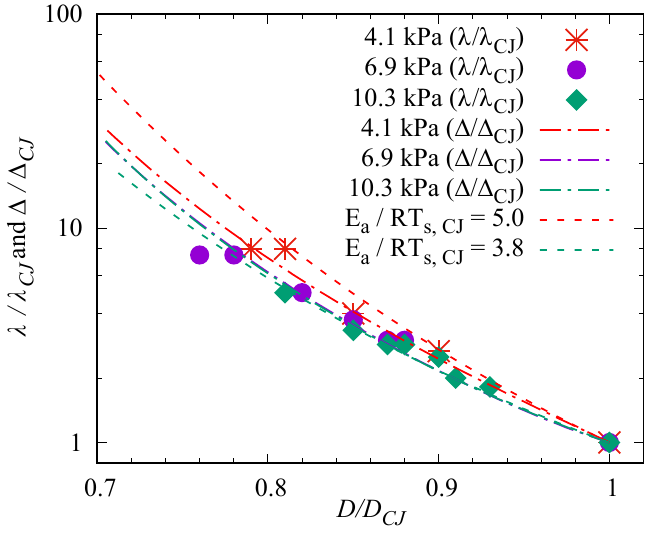}}
	\caption{The dimensionless cell size (${\lambda}/{\lambda_{CJ}}$) and characteristic induction zone length (${\Delta}/{\Delta_{CJ}}$) as a function of the velocity deficit. The symbols are from the quasi-2D simulations. The predictions (using Eq.\ (\ref{lamda-CJ-1})) with $E_a/RT_{s, CJ} = 5.0$ and $E_a/RT_{s, CJ} = 3.8$ correspond to initial pressures of 4.1 kPa and 10.3 kPa, respectively. \textcolor{black}{Note that the real-chemistry constant-volume combustion calculations ($\Delta / \Delta_{CJ}$) of 6.9 kPa (denoted by the purple dashed line) and 10.3 kPa (denoted by the green dashed line) follow almost the same curve.}} \label{Cellsize-DDCJ}  
\end{figure}

We have further obtained the dimensionless relationships in terms of the detonation cell size ($\lambda$) and the characteristic induction zone length ($\Delta$) with respect to the velocity deficit, as shown in Fig.\ \ref{Cellsize-DDCJ}. $\lambda_{CJ}$ and $\Delta_{CJ}$ represent the corresponding length scales of the ideal CJ detonation. The induction length $\Delta$ was obtained through zero-dimensional constant-volume combustion calculations with the post-shock velocity (in the shock-attached frame of reference) multiplying by the time to the peak thermicity, as proposed by Shepherd \cite{shepherd1986}. The detailed chemistry mechanism was utilized in these calculations. The excellent agreement between the ${\lambda}/{\lambda_{CJ}}$ and ${\Delta}/{\Delta_{CJ}}$ correlations suggest that the increase in cell size due to wall losses is still controlled by the increase in the induction zone length as a result of the velocity deficits.

Assuming that the cell size varies with the induction zone thickness \cite{shepherd1986} and that the ignition delay $t_{ig} \sim \textrm{exp}(E_a/RT_s)$, we can thus get
\begin{align}
\dfrac{\lambda}{\lambda_{CJ}} &\approxeq \dfrac{\Delta}{\Delta_{CJ}} = \dfrac{u'_s }{u'_{s, CJ}}\times \dfrac{t_{ig}}{t_{ig, CJ}} \notag \\
& \approxeq\left(\dfrac{u'_s}{u'_0}\right)\left(\dfrac{u'_0}{u'_{s, CJ}}\right)\textrm{exp}\left\lbrace \frac{E_a}{RT_{s, CJ}}\left[\left(\dfrac{T_{s, CJ}}{T_0}\right)\left(\dfrac{T_0}{T_{s}}\right) - 1\right] \right\rbrace \label{lamda-CJ-1}
\end{align}where $u'_s$ and $u'_0$ are the flow velocity behind and ahead of the shock in the shock-attached frame of reference, respectively, and  $T_s$ the post-shock temperature. The terms in brackets can be readily evaluated from the shock-jump equations. In the limit of strong shock and high activation energy, we can further simplify Eq.\ (\ref{lamda-CJ-1}) to
${\lambda}/{\lambda_{CJ}} \approxeq \textrm{exp}\left\lbrace \left({2E_a}/{RT_{s, CJ}}\right)\left(1 - {D}/{D_{CJ}}\right) \right\rbrace \label{lamda-CJ-2}$. It thus highlights the exponential sensitivity of the cell size and induction zone length on velocity deficit, which is controlled by the global activation energy.  This generalizes Desbordes' observations for overdriven detonations \cite{desbordes1988}. The results in Fig.\ \ref{Cellsize-DDCJ} show that the  ${\lambda}/{\lambda_{CJ}} - {D}/{D_{CJ}}$ correlations can be well predicted by the simple expression \eqref{lamda-CJ-1}. The sensitivity of cell size on velocity deficits also highlights the importance of providing the detonation speed when reporting experimentally measured cell size, since the variation can be up to an order of magnitude, even for the weakly sensitive mixtures studied here.

\section{Conclusions}
\label{conclusion}
The present study has shown that the dynamics of 2D cellular detonations in narrow channels can be well captured using a quasi-2D approach modelling the lateral boundary layer losses using Mirels' theory.  With an appropriate Mirels' constant, $K_M$, deviating by approximately a factor of 2 from the model proposed by Mirels for steady constant pressure boundary layers,  the simulations are found in excellent agreement with experiment. \textcolor{black}{Compared to directly resolving the boundary layers of detonations in the present narrow channel experiments by calculating the 3D NS equations, this novel formulation can save the computational cost by up to five orders of magnitude.}  We have also shown that the cellular cycle dynamics is also affected by the losses, which yield larger velocity fluctuations and more rapid decay rates of the lead shock.  Finally, the increase in cell size with increasing velocity deficit follows the Arrhenius dependence of ignition delay on the temperature of an equivalent steady shock, in spite of the cellular dynamics, generalizing previous observations of Desbordes for overdriven detonations in generally regular mixtures. 

\section*{Acknowledgments}
\label{Acknowledgments}
The authors wish to acknowledge the financial support from the Natural Sciences and Engineering
Research Council of Canada (NSERC) through the Discovery Grant ``Predictability of detonation wave dynamics in gases: experiment and model development", as well as the support of Compute Canada and \textcolor{black}{Core Facility for Advanced Research Computing at CWRU}.  
%\section*{References}
%% References can be added with or without bibTeX database
%% Note that the PROCI references style is considered Elsevier non-standard.
%% The original Elsevier bibliography style, elsarticle-num.bst prints paper titles as part of the references, which is different from 
%\bibliography{template.bib} %%User-specified
\bibliographystyle{elsarticle-num-PROCI}
\bibliography{references-zotero}

\end{document}